\documentclass[aps,prl,twocolumn,showpacs,floatfix]{revtex4-1} 
\usepackage{amsmath,amssymb,bm,graphicx,times}

\newcommand{\be}{\begin{equation}}
\newcommand{\ee}{\end{equation}}
\newcommand{\bea}{\begin{eqnarray}}
\newcommand{\eea}{\end{eqnarray}}
\newcommand{\bw}{\begin{widetext}}
\newcommand{\ew}{\end{widetext}}
\newcommand{\kommentar}[1]{}

\begin{document}
 
\title{Enhanced Quantum Transport in Multiplex Networks}
\author{Oliver M{\"u}lken}
\email{muelken@physik.uni-freiburg.de}
\affiliation{
Physikalisches Institut, Universit\"at Freiburg,
Hermann-Herder-Stra{\ss}e 3, 79104 Freiburg, Germany}

\date{\today} 
\begin{abstract}
Quantum transport through disordered structures is inhibited by (Anderson)
localization effects. The disorder can be either topological as in random networks or energetical as in the original Anderson model. In both cases the eigenstates of the Hamiltonian associated with the network become localized. We show how to overcome localization by network multiplexing. Here, multiple layers of random networks with the same number of nodes are stacked in such a way that in the perpendicular directions regular one-dimensional networks are formed. Depending on the ratio of the coupling within the layer and perpendicular to it, transport gets either enhanced or diminished. In particular, if the couplings are of the same order, transport gets enhanced and localization effects can be overcome. We exemplify our results by two examples: multiplexes of random networks and of one-dimensional Anderson models.
\end{abstract}
\maketitle

\begin{figure}[h]
\centerline{\includegraphics[clip=,width=0.85\columnwidth]{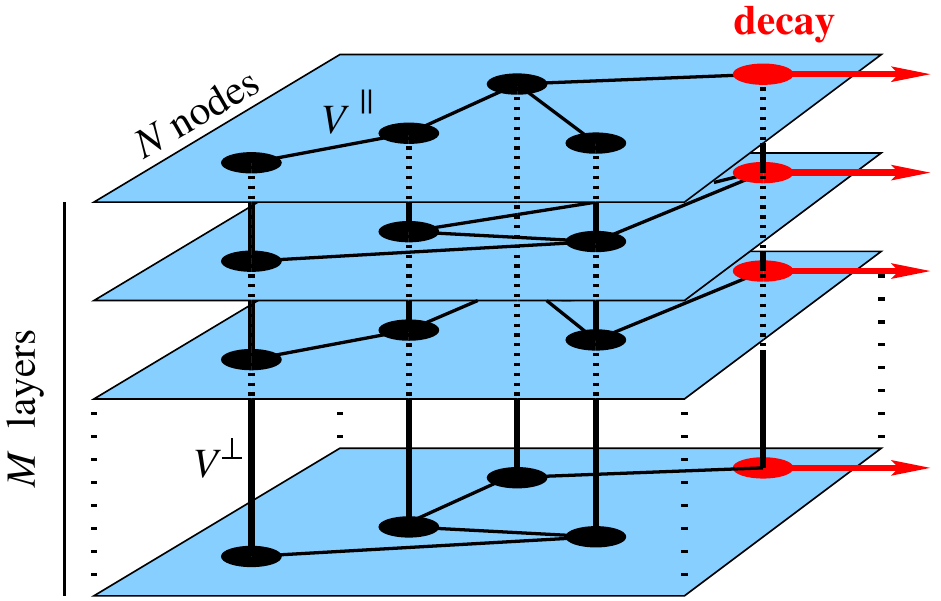}}
\caption{
(Color online) Sketch of a multiplex network with (random) networks of $N$ nodes in each of the $M$ layers.}
\label{multiplex-networks-sketch}
\end{figure}

{\sl Introduction} --
On a coarse grained scale, many physical, chemical, biological, or sociological systems can be modelled by networks of connected nodes, where every node represents one of the partners (atoms, molecules, people, etc.) interacting with each other via the bonds. Recently, there has been active research on networks of networks, i.e., on possibly different networks which are interdependent \cite{gao2011robustness,gao2012networks,kenett2014network}. A particular example are so-called multiplex networks, which are layers of interdependent networks \cite{gomez2012evolution,gomez2013diffusion,bianconi2013statistical,boccaletti2014structure}, see also Fig.~\ref{multiplex-networks-sketch}. One of the interesting aspects is the (incoherent/diffusive) dynamics of excitations within these networks \cite{gomez2013diffusion}, especially when the networks within a layer are non-regular, e.g., random or of small-world type \cite{monasson1999diffusion}.

When one is concerned with the coherent quantum dynamics of an excitation on a potential landscape given by the network, the dynamical properties can be vastly different \cite{mulken2011continuous}, while the static properties of networks can be considered to be the same also for quantum mechanical problems. There are several ways to quantify the (global/averaged) efficiency of the coherent and of the incoherent dynamics, one of which being the so-called first passage time, i.e., the time it takes for an excitation to first reach a given node of the network. If this node is equipped with a decay channel such that the excitation can leave the network, the time integral of the survival probability of the excitation defines the mean first passage time (MFPT) \cite{van1992stochastic}. This definition can be applied for both, incoherent and coherent processes, where the latter needs some adjustment, see below. For diffusive processes such as random walks, MFPTs have been calculated, e.g., for polymer reaction kinetics \cite{guerin2012non} or  confined systems \cite{benichou2014first}, see \cite{metzler2014first} for an overview of different applications.

For the quantum case, non-regular networks can lead to localization effects, e.g., if the network (in each layer) can be described by a random matrix \cite{guhr1998random} or by a one-dimensional Anderson model \cite{anderson1958absence}. These localizations prevent coherent transport, consequently leading to very large first passage times. As we are going to show, this effect can be overcome by considering multiplex networks of non-regular intra-layer networks which are regularly interconnected, see Fig.~\ref{multiplex-networks-sketch}.









{\sl Concept} --
We take the idea of multiplex networks to the quantum domain. 
To each node $k$ of the network one associates a basis vector/state $|k\rangle$, all of which form an orthonormal basis of the available state space, i.e., the Hilbert space for quantum mechanical problems \cite{mulken2011continuous}. The connectivity of the network is captured by the adjacency matrix, which has only non-zero elements if two nodes are connected by a single bond, implying that the diagonal elements are zero. The non-zero elements can be all equal, as, e.g., in random networks for which the distribution of the non-zero elements is random \cite{bollobas1982graph,monasson1999diffusion}, hence, becoming a random matrix \cite{guhr1998random}. It is also common to have non-zero elements not being equal, but rather randomly distributed in value with a regular distribution in the adjacency matrix, as, e.g., in the Anderson model \cite{anderson1958absence}. In any case, we will identify the adjacency matrix with the (discrete) Hamiltonian of the system, see below.

We are further interested in the excitation transport through such networks. By allowing for localized decay of the probability at specified nodes, we are able to monitor the global decay of the survival probability of the excitation which is a measure for the transport efficiency \cite{van1992stochastic,mulken2011continuous,mulken2007survival}: If the decay is fast, transport from an initial node to the node from which decay is possible is efficient and vice versa.

For each layer $m$ ($m=1,\dots,M$) of the multiplex network, we label the states corresponding to the nodes by $|1_m\rangle, \dots, |N_m\rangle$.
When having (localized) decay, the problem can be treated by employing an effective non-Hermitian intra-layer Hamiltonian \cite{mulken2007survival} which reads
\be 
\bm H_m^{||} = \sum_{\langle j_m k_m\rangle} V_{j_m k_m}^{||}| j_m \rangle \langle k_m| - i \Gamma_m | N_m \rangle\langle N_m|
\ee
where $V_{j_m k_m}^{||}$ is the coupling strength between pairs $\langle j_m k_m \rangle$ ($j_m,k_m=1,\dots,N_m$) of nodes which are coupled by a single bond and where $\Gamma_m$ is the decay rate from node $|N_m\rangle$ (in the following we will assume $\Gamma_m = \Gamma$ for all $m$). Therefore, the intra-layer dynamics follows for each layer $m$ from the Liouville-von Neumann equation for the density operator $\bm\rho_m(t)$ for a single layer $m$,
$\dot{\bm\rho}_m(t) = -i [ {\bm H_m^{||}}, \bm \rho_m(t)]$ 
\cite{schijven2012modeling,schijven2012avoiding}. Note, that the right-hand side of this equation has a ``non coherent'' part arising from the decay terms in $\bm H_m^{||}$.

Now, every node $j_m$ in layer $m$ is coupled to a single corresponding node $j_{m+1}$ in the adjacent layer $m+1$ with coupling strength $V_{m,m+1}^{\perp}$. This creates chains of nodes in the direction perpendicular to the layers. Then, the inter-layer Hamiltonian is given by
\be
\bm H_j^{\perp} = \sum_{m=1}^{M-1} V_{m,m+1}^{\perp} | j_m \rangle \langle j_{m*1}| .
\ee
The full dynamics for the density operator $\bm\rho(t)$ of the whole multiplex network is then described by
$\dot{\bm\rho}(t) = -i [\bm H, \bm\rho(t)]$
with $\bm H = \sum_{m=1}^M \bm H_m^{||} + \sum_{j=1}^N \bm H_j^{\perp}$. In total there are $M$ nodes from which an excitation can decay, with corresponding states $|N_m\rangle$ with $m=1,\dots,M$.

In order to study the transport through these networks, we assume initial conditions which are localized at any node of the whole multiplex network such that the transition probability from node $k$ to node $j$ reads $\pi_{jk}(t) = \langle j|\bm\rho(t)|j\rangle$ with $\bm\rho(0)=|k\rangle\langle k|$. The problem can be properly solved by diagonalization of $\bm H$. In most cases, the
eigenvalues $E_n = \epsilon_n + i \gamma_n$ ($n=1,\dots,NM$) are complex with
$\epsilon_n\in\mathbb R$ and $\gamma_n \in \mathbb R_0^+$.
By then averaging over all possible initial nodes $k$ and all possible final nodes $j$ we obtain a global picture of the decay process, i.e., of the transport efficiency. Thus, the mean survival probability 
is given by \cite{mulken2011continuous,mulken2007survival}:
\be
\Pi(t) \equiv \frac{1}{NM} \sum_{k,j=1}^{N M} \pi_{jk}(t) = \frac{1}{NM} \sum_{n=1}^{NM} \exp(-2\gamma_n t).
\ee
In the limit when $\Gamma \ll 1$, the imaginary parts are related to the eigenstates of the network without decay, namely 
$\gamma_n = \Gamma \sum_m |\langle N_m | \Phi_n^{(0)}\rangle|^2$, where
$| \Phi_n^{(0)}\rangle$ are the eigenstate of $\bm H^{(0)} = \sum_m \bm H^{|| (0)}_m
+ \sum_j \bm H_j^{\perp}$, where $H^{|| (0)}_m$ is the hermitian part of $H^{||}_m$ \cite{mulken2011continuous}.

{\sl General Results} --
Now, since not necessarily all values of the $\gamma_n$'s need to be larger than zero, the mean survival probability can decay to a finite value given by
\be
\lim_{t\to\infty} \Pi(t) \equiv \Pi_\infty = \frac{1}{NM} \sum\limits_{\{n|\gamma_n=0\}} 1 \equiv \frac{{\cal N}_0}{NM},
\ee
where ${\cal N}_0$ is the number of vanishing values of the $\gamma_n$'s.
Therefore, and in analogy to the classical case \cite{van1992stochastic}, we now define the quantum analog of the global
mean first passage time (gMFPT) as 
\be\label{gmfpt}
\tau \equiv \int\limits_0^\infty dt \ \big[\Pi(t) - \Pi_\infty\big] 
=  
\frac{1}{NM} \sum\limits_{\{n|\gamma_n\neq 0\}} \frac{1}{2\gamma_n}.
\ee
Clearly, Eq.~(\ref{gmfpt}) depends on ${\cal N}_0$ as well as on $N$ and on $M$. In the following we will keep $N$ constant while varying $M$. 

Since the intra-layer networks we consider are random networks we will further analyze the 
ensemble averages over $R$ realizations of the multiplex network, namely $\langle\Pi_\infty\rangle_R = \sum_{r=1}^R \Pi_\infty^{[r]}$ and  $\langle \tau \rangle_R = \sum_{r=1}^R \tau^{[r]}$, where the superscript $[r]$ denotes the $r$th realization.
In order to obtain analytical approximations/bounds, we calculate 
the quantum analog of the gMFPT
from the ensemble averaged $\gamma_n$'s, $\langle\gamma_n\rangle_R = \sum_{r=1}^R \gamma_n^{[r]}$, i.e. we define 
\be
\overline{\tau} \equiv \frac{1}{NM} \sum_{\{n|\langle\gamma_n\rangle_R\neq 0\}}
\frac{1}{2 \langle\gamma_n\rangle_R}.
\ee
Jensen's inequality guarantees that 
$\overline{\tau}\leq \langle \tau \rangle_R$,
see paragraph 12.41 of \cite{gradshteyn}.
As will be shown in what follows, the properties of $\overline{\tau}$ and $\langle \tau\rangle_R$, 
as well as of 
$\langle{\cal N}_0\rangle_R$ crucially depend on the number of layers and on the ratio $V^{||}/V^{\perp}$.
For the two limits $V^{\perp}\gg V^{||}$ and $V^{\perp}\ll V^{||}$ perturbation theory yields that $\overline{\tau}$ and also $\langle \tau\rangle_R$ decrease with increasing $M$:

(i)
For  $V^{\perp}\gg V^{||}$, the total networks consists of $N$ chain-like networks, each of length $M$, which are only weakly coupled within each layer. This implies that approximately only those states $|\Phi^{(0)}\rangle$ belonging to the chain which contains the nodes with decay have non vanishing overlap with the states $|N_m\rangle$. Therefore, for $V^{||}/V^{\perp}\to 0$ there will be only $M$ values $\gamma_n\neq 0$ entering $\tau$ and consequently also only $M$ averaged values $\langle\gamma_n\rangle_R\neq 0$ entering $\overline{\tau}$. Further, one finds ${\cal N}_0 \sim (N-1)M$.
The eigenstates of each chain are Bloch-like states, i.e., they are states of the form $(1/\sqrt{M}) \sum_m c_m |N_m\rangle$, where $c_m$ is a complex number lying on the unit circle \cite{kittel}. This then leads to
$ \tau \sim (1/2N\Gamma) \sum_m 1/|c_m|^2$,
with $c_m = 1$ for Bloch-like states, such that $\tau\sim 1/(2\Gamma N)$ becomes independent of $M$. 
This result also holds in the ensemble averages, both, for $\overline{\tau}$ and for $\langle \tau \rangle_R$.

(ii)
In the other limit where $V^{\perp}\ll V^{||}$, the total multiplex network consists of very weakly coupled layers. Thus, the dynamics is mainly dominated by the dynamics within a single layer and in addition by weak tranport perpendicular to this layer. 
In this case, however, we have to consider the ensemble averages since the details of each realization of a random network can vary (strongly). The (average) eigenstates of random networks are exponentially localized \cite{monasson1999diffusion}, such that a significant overlap of these states with $|N_m\rangle$, occurs only for a number of order ${\cal O}(1)$ of these states. When assuming for simplicity that all the other states have zero overlap, the sum in $\overline{\tau}$ will only run over those number of states with overlap which for the whole multiplex network is of order ${\cal O}(M)$, again implying that ${\cal N}_0 \sim (N-1)M$. Consequently, also here $\overline\tau$ 
is independent of $M$. As it turns out, the same holds for the ensemble average $\langle \tau \rangle_R$.

Having, in the two limits (i) and (ii), for an increasing number of layers, $M$, a linear increase in the number of vanishing values of the $\gamma_n$'s and constant values for the quantum analogs of the gMFPTs does not imply that this is true for all values of $V^{||}/V^{\perp}$. As our calculations show, one finds a {\it decrease} of all considered quantities with $M$ when $V^{\perp}\approx V^{||}$, i.e., when no perturbation limit can be applied.

The two limits (i) and (ii) show that the minimal number of values $\gamma_n=0$ is given by ${\cal N}_0 \sim (N-1)M$. For fixed $M$, any change in the couplings leading away from these limits will lead to an increase of this number. This is best seen on the basis of the eigenstates $|\Phi_N^{(0)}\rangle$: A change in the ratio $V^{||}/V^{\perp}$ away from the limits (i) and (ii) results in eigenstates which are distributed over larger areas of the multiplex network. Consequently, the number of values $\gamma_n = \Gamma \sum_m |\langle N_m | \Phi_n^{(0)}\rangle|^2 =0$ will decrease and more probability can leak out of the system, leading to a lowering of $\Pi_\infty$. Now, increasing $M$ will result in a further decrease of ${\cal N}_0$ and of $\Pi_\infty$.

Similar arguments hold for $\tau$: Since the eigenstates become ``broader'' with in-/decreasing the ratio $V^{||}/V^{\perp}$ from the limit (i)/(ii), the overlap of most eigenstates with $|N_m\rangle$ becomes larger, leading to a larger value of the corresponding $\gamma_n$. This happens on the cost of the overlap of other eigenstates with $|N_m\rangle$, since $\sum_n |\langle N_m | \Phi_n^{(0)}\rangle|^2 = 1$ holds.
Suppose that there are only two states with overlap such that, say, $a_j \equiv |\langle N_m | \Phi_j^{(0)}\rangle|^2$ with $j=1,2$, for which $a_1>a_2$ and $a_1+a_2=1$. After a change in $V^{||}/V^{\perp}$, we assume to have new states with $\tilde a_1=a_1-\varepsilon$ and $\tilde a_2 = a_2 -\varepsilon$, where $1\gg\varepsilon>0$, such that also $\tilde a_1 + \tilde a_2 =1$. For $\tau$ we need $1/\tilde a_1 + 1/\tilde a_2$, which after expanding in $\varepsilon$ becomes $1/a_1 + 1/a_2 - \varepsilon (1-2a_2)/[a_2^2(1-a_2)] < 1/a_1 + 1/a_2$.
As a result, these change in the values of the $\gamma_n$'s leads to a {\it smaller} value of $\tau$. Also here, increasing $M$ leads to a further decrease of $\tau$.


{\sl Examples} --
We will exemplify the above results by two variants of multiplex networks. In order to minimize the numerical effort but still obtain reasonable results, we take for both examples $N=10$, a trapping rate of $\Gamma=0.001 \cdot V^{||}$,  and the number of realizations $R=200$.

First, we consider multiplex networks where each layer consists of a random network of given size $N$, as sketched in Fig.~\ref{multiplex-networks-sketch}. The maximal number of bonds in each network is $N(N-1)/2$, out of which we take a fraction of $1/2$ to be placed at random between the bonds, such that maximally one bond connects two nodes. We assume the same coupling strength $V^{||}=1$ for any pair of nodes which is connected by a single bond. In the second example, we take the network in each layer to be a one-dimensional Anderson model with off-diagonal disorder, again with $N=10$. Here the disordered couplings $V_{j_m,k_m}^{||} = V^{||} + \Delta_{j_m,k_m}$ are drawn from a normal distribution centered around the value $V^{||}=1$. In both cases $V^\perp$ will be given in units of $V^{||}$.

Since all quantities we consider depend on the values of the $\gamma_n$'s, in particular on those values which are either exactly zero or strictly larger than zero, it makes it numerically difficult to determine whether a value is truly zero or not. This can cause large deviations in, say, the values of $\tau$, see Eq.~(\ref{gmfpt}). Therefore, we use a cut-off below which we assume the $\gamma_n$'s to be zero, i.e., we neglect long-time tails of $\Pi(t)$ and also slightly raise the value of $\Pi_\infty$. We have checked different values of the cut-off and found a value of $10^{-6}$ a reasonable choice, because the fluctuations in the considered quantities were rather small, see the numerical results below. 
We note that choosing too large a cut-off could deminish the differences discussed above. 
 

\begin{figure}[h]
\centerline{\includegraphics[clip=,width=\columnwidth]{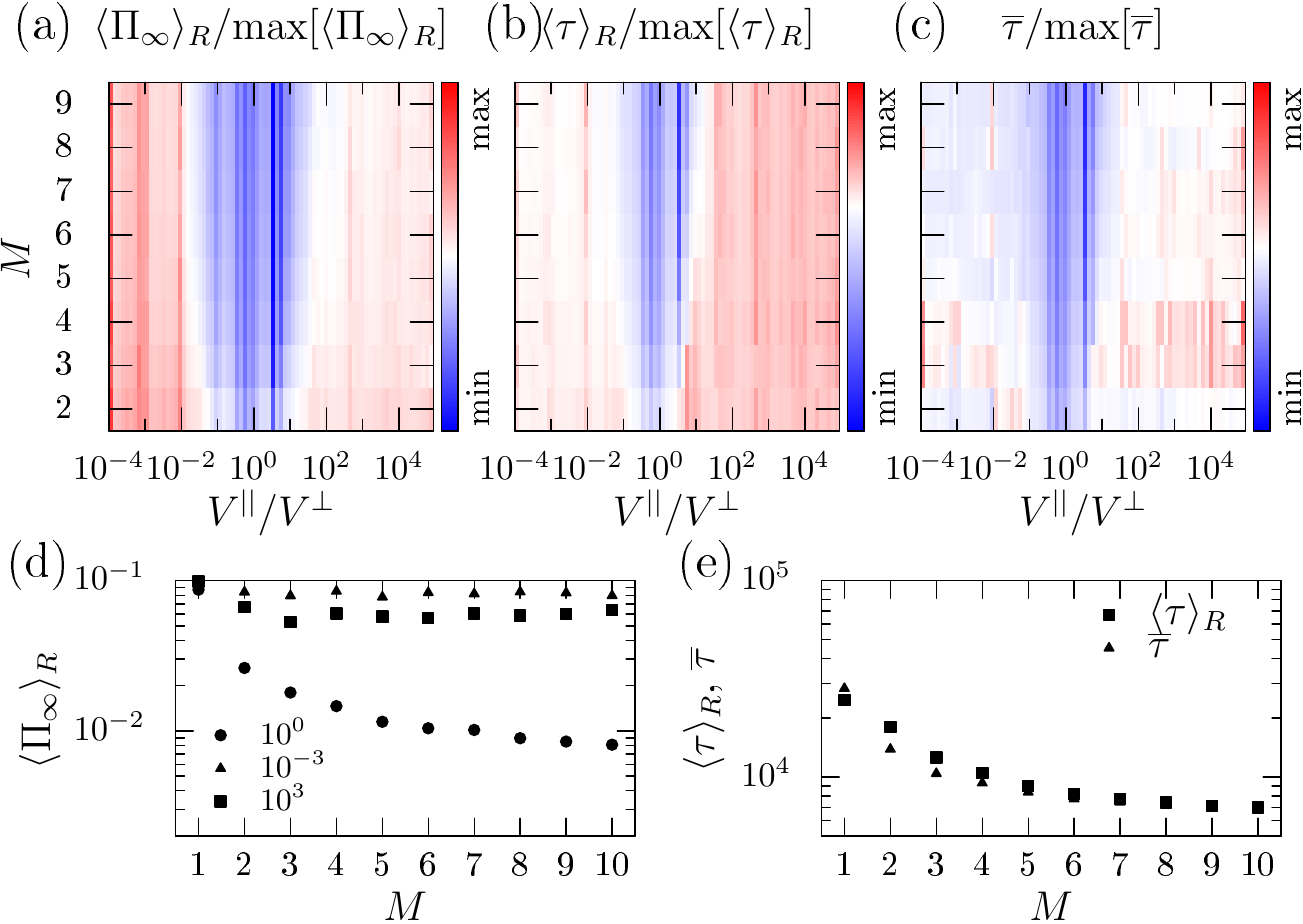}}
\caption{
(Color online) Multiplex networks with each layer being a random network of size $N=10$ with a fraction of $1/2$ of all possible bonds present. (a)-(c) show contour plots of $\langle\Pi_\infty\rangle_R$, $\langle \tau \rangle_R$, and 
$\overline\tau$, respectively, for different ratios of the couplings $V^{||}/V^{\perp}$ and different numbers of layers, $M$. Note that all plots are scaled by the respective maximal values. There is a clear minimum for all three quantities at value $V^{||}/V^{\perp}\approx 1$. The lower panels illustrate the $M$-dependence (d) of $\langle\Pi_\infty\rangle_R$ for different values of $V^{||}/V^{\perp}$ and (e) of $\langle \tau \rangle_R$ and $\overline\tau$ for $V^{||}/V^{\perp}=1$. Especially for $\langle \tau \rangle_R$ there is a drop by about one order of magnitude from $M=1$ to $M=10$.}
\label{plot-random}
\end{figure}

For multiplexes of random networks, Figs.~\ref{plot-random}(a)-(b) show $\langle\Pi_\infty\rangle_R$, $\langle \tau \rangle_R$, and 
$\overline\tau$, respectively, for different ratios $V^{||}/V^{\perp}$ and different $M$ and rescaled to their respective maximal values. The first thing to notice is that in all three cases there is a pronounced minimum for ratios $V^{||}/V^{\perp}\approx 1$, which decreases with increasing $M$, thus corroborating the general statements given above. This is further illustrated in Fig.~\ref{plot-random}(d) where the $M$-dependence of $\langle\Pi_\infty\rangle_R$ for different values of $V^{||}/V^{\perp}$ is shown and in Fig.~\ref{plot-random}(e) where the $M$-dependence of $\langle \tau \rangle_R$ and $\overline\tau$ for $V^{||}/V^{\perp}=1$ is displayed. In particular, $\langle \tau \rangle_R$ shows a drop by about one order of magnitude from $M=1$ to $M=10$. This indicates that one can overcome localization effects by a suitable choice of multiplexing several (random) networks.
Moreover, the results also confirm that in the two limits (i) and (ii) all three quantitites do not change with $M$ for $M\geq 2$. 
For all ratios $V^{||}/V^{\perp}$ , there is a change in all quantitites when going from $M=1$ to $M=2$, see Fig~\ref{plot-random}(d) and (e). In addition, we find the trivial result that for $M=1$ (only one layer, i.e., no multiplex) the three quantities do not change with the ratio $V^{||}/V^{\perp}$, see Fig~\ref{plot-random}(d). 

\begin{figure}[h]
\centerline{\includegraphics[clip=,width=\columnwidth]{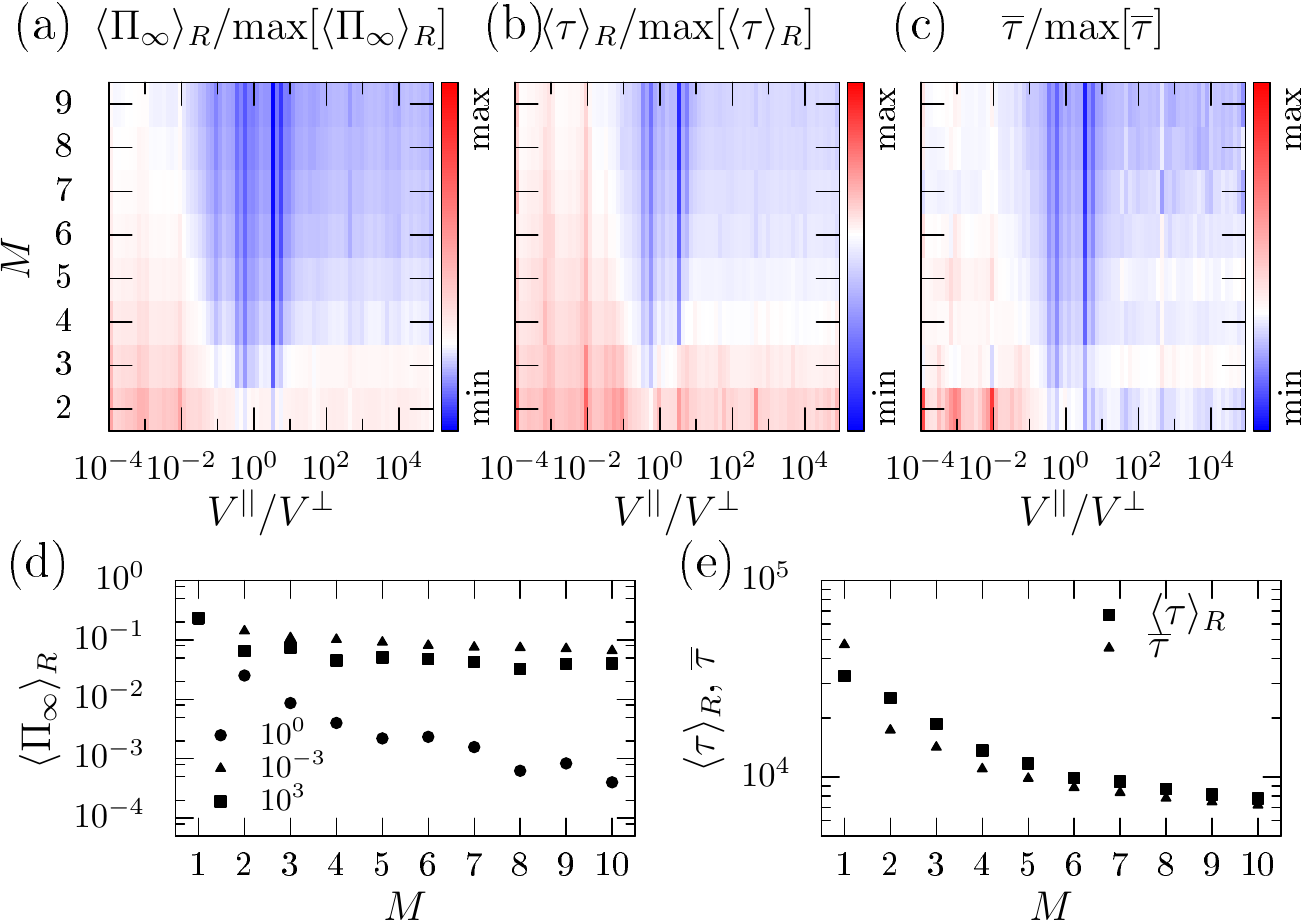}}
\caption{
(Color online) Multiplex networks with each layer being an Anderson model of $N=10$ nodes with off-diagonal disorder drawn from a normal distribution with width $\sigma = 0.5 V^{||}$. See Fig.~\ref{plot-random} for description of the different panels.}
\label{plot-anderson}
\end{figure}

For multiplexes of Anderson models, Fig.~\ref{plot-anderson} shows the same quantities as Fig.~\ref{plot-random} for slightly different parameter ranges of $V^{||}/V^{\perp}$ and $M$. Given similar features of the eigenstates $|\Phi^{(0)}\rangle$ of random networks and of the Anderson model, we find similar results as for the multiplexes of random networks, see Figs.~\ref{plot-anderson}(a)-(c). Most importantly, also here there is a pronounced minimum at values $V^{||}/V^{\perp}\approx 1$, where $\langle\Pi_\infty\rangle_R$ decreases by about two orders of magnitude from $M=1$ to $M=6$, see Fig.~\ref{plot-anderson}(d). The quantum analog of the gMFPT, $\langle \tau \rangle_R$, decreases in the same interval of $M$ by about one order of magnitude. Again, suitable multiplexing of several Anderson model overcomes localization and the overall transport through the whole network becomes (much) more efficient.



{\sl Conclusions} --
We have shown that the coherent quantum transport on non-regular networks which show localization can be enhanced arranging these networks in layers which are regularly interconnected. The quantum analog of the global mean first passage time allows to define a global measure for the transport efficiency, which depends on the 
ratio between the intra- and the inter-network coupling strengths, $V^{||}/V^{\perp}$ and the number of layers, $M$.
The ratio $V^{||}/V^{\perp}$ determines how large the enhancement will be: While in the limits $V^{||}/V^{\perp}\ll1$ and $V^{||}/V^{\perp}\gg1$ there is only very little enhancement,  one can achieve a significant enhancemnt for $V^{||}/V^{\perp}\approx 1$. Our general analytical results are corroborated by numerical calculations for two examples, multiplex networks of intra-layer random networks and of intra-layer one-dimensional Anderson models. 
We believe that these model systems can also be realized experimentally, say, using wave-guide arrays for which, e.g., Anderson localization has been realized \cite{lahini2008anderson,martin2011anderson,segev2013anderson}. One can also envision that multiplex networks could be used to improve signal transfer.



The author thanks Alexander Blumen for many fruitful discussions and valuable comments.
Support from the Deutsche For\-schungs\-ge\-mein\-schaft (DFG Grant No. MU2925/1-1) is gratefully acknowledged.

%

\begin{thebibliography}{24}%
\makeatletter
\providecommand \@ifxundefined [1]{%
 \@ifx{#1\undefined}
}%
\providecommand \@ifnum [1]{%
 \ifnum #1\expandafter \@firstoftwo
 \else \expandafter \@secondoftwo
 \fi
}%
\providecommand \@ifx [1]{%
 \ifx #1\expandafter \@firstoftwo
 \else \expandafter \@secondoftwo
 \fi
}%
\providecommand \natexlab [1]{#1}%
\providecommand \enquote  [1]{``#1''}%
\providecommand \bibnamefont  [1]{#1}%
\providecommand \bibfnamefont [1]{#1}%
\providecommand \citenamefont [1]{#1}%
\providecommand \href@noop [0]{\@secondoftwo}%
\providecommand \href [0]{\begingroup \@sanitize@url \@href}%
\providecommand \@href[1]{\@@startlink{#1}\@@href}%
\providecommand \@@href[1]{\endgroup#1\@@endlink}%
\providecommand \@sanitize@url [0]{\catcode `\\12\catcode `\$12\catcode
  `\&12\catcode `\#12\catcode `\^12\catcode `\_12\catcode `\%12\relax}%
\providecommand \@@startlink[1]{}%
\providecommand \@@endlink[0]{}%
\providecommand \url  [0]{\begingroup\@sanitize@url \@url }%
\providecommand \@url [1]{\endgroup\@href {#1}{\urlprefix }}%
\providecommand \urlprefix  [0]{URL }%
\providecommand \Eprint [0]{\href }%
\providecommand \doibase [0]{http://dx.doi.org/}%
\providecommand \selectlanguage [0]{\@gobble}%
\providecommand \bibinfo  [0]{\@secondoftwo}%
\providecommand \bibfield  [0]{\@secondoftwo}%
\providecommand \translation [1]{[#1]}%
\providecommand \BibitemOpen [0]{}%
\providecommand \bibitemStop [0]{}%
\providecommand \bibitemNoStop [0]{.\EOS\space}%
\providecommand \EOS [0]{\spacefactor3000\relax}%
\providecommand \BibitemShut  [1]{\csname bibitem#1\endcsname}%
\let\auto@bib@innerbib\@empty
\bibitem [{\citenamefont {Gao}\ \emph {et~al.}(2011)\citenamefont {Gao},
  \citenamefont {Buldyrev}, \citenamefont {Havlin},\ and\ \citenamefont
  {Stanley}}]{gao2011robustness}%
  \BibitemOpen
  \bibfield  {author} {\bibinfo {author} {\bibfnamefont {J.}~\bibnamefont
  {Gao}}, \bibinfo {author} {\bibfnamefont {S.~V.}\ \bibnamefont {Buldyrev}},
  \bibinfo {author} {\bibfnamefont {S.}~\bibnamefont {Havlin}}, \ and\ \bibinfo
  {author} {\bibfnamefont {H.~E.}\ \bibnamefont {Stanley}},\ }\href@noop {}
  {\bibfield  {journal} {\bibinfo  {journal} {Phys. Rev. Lett.}\ }\textbf
  {\bibinfo {volume} {107}},\ \bibinfo {pages} {195701} (\bibinfo {year}
  {2011})}\BibitemShut {NoStop}%
\bibitem [{\citenamefont {Gao}\ \emph {et~al.}(2012)\citenamefont {Gao},
  \citenamefont {Buldyrev}, \citenamefont {Stanley},\ and\ \citenamefont
  {Havlin}}]{gao2012networks}%
  \BibitemOpen
  \bibfield  {author} {\bibinfo {author} {\bibfnamefont {J.}~\bibnamefont
  {Gao}}, \bibinfo {author} {\bibfnamefont {S.~V.}\ \bibnamefont {Buldyrev}},
  \bibinfo {author} {\bibfnamefont {H.~E.}\ \bibnamefont {Stanley}}, \ and\
  \bibinfo {author} {\bibfnamefont {S.}~\bibnamefont {Havlin}},\ }\href@noop {}
  {\bibfield  {journal} {\bibinfo  {journal} {Nature Physics}\ }\textbf
  {\bibinfo {volume} {8}},\ \bibinfo {pages} {40} (\bibinfo {year}
  {2012})}\BibitemShut {NoStop}%
\bibitem [{\citenamefont {Kenett}\ \emph {et~al.}(2014)\citenamefont {Kenett},
  \citenamefont {Gao}, \citenamefont {Huang}, \citenamefont {Shao},
  \citenamefont {Vodenska}, \citenamefont {Buldyrev}, \citenamefont {Paul},
  \citenamefont {Stanley},\ and\ \citenamefont {Havlin}}]{kenett2014network}%
  \BibitemOpen
  \bibfield  {author} {\bibinfo {author} {\bibfnamefont {D.~Y.}\ \bibnamefont
  {Kenett}}, \bibinfo {author} {\bibfnamefont {J.}~\bibnamefont {Gao}},
  \bibinfo {author} {\bibfnamefont {X.}~\bibnamefont {Huang}}, \bibinfo
  {author} {\bibfnamefont {S.}~\bibnamefont {Shao}}, \bibinfo {author}
  {\bibfnamefont {I.}~\bibnamefont {Vodenska}}, \bibinfo {author}
  {\bibfnamefont {S.~V.}\ \bibnamefont {Buldyrev}}, \bibinfo {author}
  {\bibfnamefont {G.}~\bibnamefont {Paul}}, \bibinfo {author} {\bibfnamefont
  {H.~E.}\ \bibnamefont {Stanley}}, \ and\ \bibinfo {author} {\bibfnamefont
  {S.}~\bibnamefont {Havlin}},\ }in\ \href@noop {} {\emph {\bibinfo {booktitle}
  {Networks of Networks: The Last Frontier of Complexity}}}\ (\bibinfo
  {publisher} {Springer},\ \bibinfo {year} {2014})\ p.~\bibinfo {pages}
  {3}\BibitemShut {NoStop}%
\bibitem [{\citenamefont {G{\'o}mez-Gardenes}\ \emph
  {et~al.}(2012)\citenamefont {G{\'o}mez-Gardenes}, \citenamefont {Reinares},
  \citenamefont {Arenas},\ and\ \citenamefont
  {Flor{\'\i}a}}]{gomez2012evolution}%
  \BibitemOpen
  \bibfield  {author} {\bibinfo {author} {\bibfnamefont {J.}~\bibnamefont
  {G{\'o}mez-Gardenes}}, \bibinfo {author} {\bibfnamefont {I.}~\bibnamefont
  {Reinares}}, \bibinfo {author} {\bibfnamefont {A.}~\bibnamefont {Arenas}}, \
  and\ \bibinfo {author} {\bibfnamefont {L.~M.}\ \bibnamefont {Flor{\'\i}a}},\
  }\href@noop {} {\bibfield  {journal} {\bibinfo  {journal} {Scientific
  Reports}\ }\textbf {\bibinfo {volume} {2}} (\bibinfo {year}
  {2012})}\BibitemShut {NoStop}%
\bibitem [{\citenamefont {Gomez}\ \emph {et~al.}(2013)\citenamefont {Gomez},
  \citenamefont {Diaz-Guilera}, \citenamefont {Gomez-Garde{\~n}es},
  \citenamefont {Perez-Vicente}, \citenamefont {Moreno},\ and\ \citenamefont
  {Arenas}}]{gomez2013diffusion}%
  \BibitemOpen
  \bibfield  {author} {\bibinfo {author} {\bibfnamefont {S.}~\bibnamefont
  {Gomez}}, \bibinfo {author} {\bibfnamefont {A.}~\bibnamefont {Diaz-Guilera}},
  \bibinfo {author} {\bibfnamefont {J.}~\bibnamefont {Gomez-Garde{\~n}es}},
  \bibinfo {author} {\bibfnamefont {C.~J.}\ \bibnamefont {Perez-Vicente}},
  \bibinfo {author} {\bibfnamefont {Y.}~\bibnamefont {Moreno}}, \ and\ \bibinfo
  {author} {\bibfnamefont {A.}~\bibnamefont {Arenas}},\ }\href@noop {}
  {\bibfield  {journal} {\bibinfo  {journal} {Phys. Rev. Lett.}\ }\textbf
  {\bibinfo {volume} {110}},\ \bibinfo {pages} {028701} (\bibinfo {year}
  {2013})}\BibitemShut {NoStop}%
\bibitem [{\citenamefont {Bianconi}(2013)}]{bianconi2013statistical}%
  \BibitemOpen
  \bibfield  {author} {\bibinfo {author} {\bibfnamefont {G.}~\bibnamefont
  {Bianconi}},\ }\href@noop {} {\bibfield  {journal} {\bibinfo  {journal}
  {Phys. Rev. E}\ }\textbf {\bibinfo {volume} {87}},\ \bibinfo {pages} {062806}
  (\bibinfo {year} {2013})}\BibitemShut {NoStop}%
\bibitem [{\citenamefont {Boccaletti}\ \emph {et~al.}(2014)\citenamefont
  {Boccaletti}, \citenamefont {Bianconi}, \citenamefont {Criado}, \citenamefont
  {Del~Genio}, \citenamefont {G{\'o}mez-Garde{\~n}es}, \citenamefont {Romance},
  \citenamefont {Sendina-Nadal}, \citenamefont {Wang},\ and\ \citenamefont
  {Zanin}}]{boccaletti2014structure}%
  \BibitemOpen
  \bibfield  {author} {\bibinfo {author} {\bibfnamefont {S.}~\bibnamefont
  {Boccaletti}}, \bibinfo {author} {\bibfnamefont {G.}~\bibnamefont
  {Bianconi}}, \bibinfo {author} {\bibfnamefont {R.}~\bibnamefont {Criado}},
  \bibinfo {author} {\bibfnamefont {C.}~\bibnamefont {Del~Genio}}, \bibinfo
  {author} {\bibfnamefont {J.}~\bibnamefont {G{\'o}mez-Garde{\~n}es}}, \bibinfo
  {author} {\bibfnamefont {M.}~\bibnamefont {Romance}}, \bibinfo {author}
  {\bibfnamefont {I.}~\bibnamefont {Sendina-Nadal}}, \bibinfo {author}
  {\bibfnamefont {Z.}~\bibnamefont {Wang}}, \ and\ \bibinfo {author}
  {\bibfnamefont {M.}~\bibnamefont {Zanin}},\ }\href@noop {} {\bibfield
  {journal} {\bibinfo  {journal} {Phys. Rep.}\ }\textbf {\bibinfo {volume}
  {544}},\ \bibinfo {pages} {1} (\bibinfo {year} {2014})}\BibitemShut {NoStop}%
\bibitem [{\citenamefont {Monasson}(1999)}]{monasson1999diffusion}%
  \BibitemOpen
  \bibfield  {author} {\bibinfo {author} {\bibfnamefont {R.}~\bibnamefont
  {Monasson}},\ }\href@noop {} {\bibfield  {journal} {\bibinfo  {journal} {Eur.
  Phys. J. B}\ }\textbf {\bibinfo {volume} {12}},\ \bibinfo {pages} {555}
  (\bibinfo {year} {1999})}\BibitemShut {NoStop}%
\bibitem [{\citenamefont {M{\"u}lken}\ and\ \citenamefont
  {Blumen}(2011)}]{mulken2011continuous}%
  \BibitemOpen
  \bibfield  {author} {\bibinfo {author} {\bibfnamefont {O.}~\bibnamefont
  {M{\"u}lken}}\ and\ \bibinfo {author} {\bibfnamefont {A.}~\bibnamefont
  {Blumen}},\ }\href@noop {} {\bibfield  {journal} {\bibinfo  {journal} {Phys.
  Rep.}\ }\textbf {\bibinfo {volume} {502}},\ \bibinfo {pages} {37} (\bibinfo
  {year} {2011})}\BibitemShut {NoStop}%
\bibitem [{\citenamefont {Van~Kampen}(1992)}]{van1992stochastic}%
  \BibitemOpen
  \bibfield  {author} {\bibinfo {author} {\bibfnamefont {N.~G.}\ \bibnamefont
  {Van~Kampen}},\ }\href@noop {} {\emph {\bibinfo {title} {Stochastic processes
  in physics and chemistry}}},\ Vol.~\bibinfo {volume} {1}\ (\bibinfo
  {publisher} {Elsevier},\ \bibinfo {year} {1992})\BibitemShut {NoStop}%
\bibitem [{\citenamefont {Gu{\'e}rin}\ \emph {et~al.}(2012)\citenamefont
  {Gu{\'e}rin}, \citenamefont {B{\'e}nichou},\ and\ \citenamefont
  {Voituriez}}]{guerin2012non}%
  \BibitemOpen
  \bibfield  {author} {\bibinfo {author} {\bibfnamefont {T.}~\bibnamefont
  {Gu{\'e}rin}}, \bibinfo {author} {\bibfnamefont {O.}~\bibnamefont
  {B{\'e}nichou}}, \ and\ \bibinfo {author} {\bibfnamefont {R.}~\bibnamefont
  {Voituriez}},\ }\href@noop {} {\bibfield  {journal} {\bibinfo  {journal}
  {Nature Chemistry}\ }\textbf {\bibinfo {volume} {4}},\ \bibinfo {pages} {568}
  (\bibinfo {year} {2012})}\BibitemShut {NoStop}%
\bibitem [{\citenamefont {B{\'e}nichou}\ and\ \citenamefont
  {Voituriez}(2014)}]{benichou2014first}%
  \BibitemOpen
  \bibfield  {author} {\bibinfo {author} {\bibfnamefont {O.}~\bibnamefont
  {B{\'e}nichou}}\ and\ \bibinfo {author} {\bibfnamefont {R.}~\bibnamefont
  {Voituriez}},\ }\href@noop {} {\bibfield  {journal} {\bibinfo  {journal}
  {Phys. Rep.}\ }\textbf {\bibinfo {volume} {539}},\ \bibinfo {pages} {225}
  (\bibinfo {year} {2014})}\BibitemShut {NoStop}%
\bibitem [{\citenamefont {Metzler}\ \emph {et~al.}(2014)\citenamefont
  {Metzler}, \citenamefont {Oshanin},\ and\ \citenamefont
  {Redner}}]{metzler2014first}%
  \BibitemOpen
  \bibfield  {author} {\bibinfo {author} {\bibfnamefont {R.}~\bibnamefont
  {Metzler}}, \bibinfo {author} {\bibfnamefont {G.}~\bibnamefont {Oshanin}}, \
  and\ \bibinfo {author} {\bibfnamefont {S.}~\bibnamefont {Redner}},\
  }\href@noop {} {\emph {\bibinfo {title} {First-Passage Phenomena and Their
  Applications}}}\ (\bibinfo  {publisher} {World Scientific},\ \bibinfo {year}
  {2014})\BibitemShut {NoStop}%
\bibitem [{\citenamefont {Guhr}\ \emph {et~al.}(1998)\citenamefont {Guhr},
  \citenamefont {M{\"u}ller-Groeling},\ and\ \citenamefont
  {Weidenm{\"u}ller}}]{guhr1998random}%
  \BibitemOpen
  \bibfield  {author} {\bibinfo {author} {\bibfnamefont {T.}~\bibnamefont
  {Guhr}}, \bibinfo {author} {\bibfnamefont {A.}~\bibnamefont
  {M{\"u}ller-Groeling}}, \ and\ \bibinfo {author} {\bibfnamefont {H.~A.}\
  \bibnamefont {Weidenm{\"u}ller}},\ }\href@noop {} {\bibfield  {journal}
  {\bibinfo  {journal} {Phys. Rep.}\ }\textbf {\bibinfo {volume} {299}},\
  \bibinfo {pages} {189} (\bibinfo {year} {1998})}\BibitemShut {NoStop}%
\bibitem [{\citenamefont {Anderson}(1958)}]{anderson1958absence}%
  \BibitemOpen
  \bibfield  {author} {\bibinfo {author} {\bibfnamefont {P.~W.}\ \bibnamefont
  {Anderson}},\ }\href@noop {} {\bibfield  {journal} {\bibinfo  {journal}
  {Phys. Rev.}\ }\textbf {\bibinfo {volume} {109}},\ \bibinfo {pages} {1492}
  (\bibinfo {year} {1958})}\BibitemShut {NoStop}%
\bibitem [{\citenamefont {Bollob{\'a}s}(1982)}]{bollobas1982graph}%
  \BibitemOpen
  \bibfield  {author} {\bibinfo {author} {\bibfnamefont {B.}~\bibnamefont
  {Bollob{\'a}s}},\ }\href@noop {} {\emph {\bibinfo {title} {Graph theory}}}\
  (\bibinfo  {publisher} {Elsevier},\ \bibinfo {year} {1982})\BibitemShut
  {NoStop}%
\bibitem [{\citenamefont {M{\"u}lken}\ \emph {et~al.}(2007)\citenamefont
  {M{\"u}lken}, \citenamefont {Blumen}, \citenamefont {Amthor}, \citenamefont
  {Giese}, \citenamefont {Reetz-Lamour},\ and\ \citenamefont
  {Weidem{\"u}ller}}]{mulken2007survival}%
  \BibitemOpen
  \bibfield  {author} {\bibinfo {author} {\bibfnamefont {O.}~\bibnamefont
  {M{\"u}lken}}, \bibinfo {author} {\bibfnamefont {A.}~\bibnamefont {Blumen}},
  \bibinfo {author} {\bibfnamefont {T.}~\bibnamefont {Amthor}}, \bibinfo
  {author} {\bibfnamefont {C.}~\bibnamefont {Giese}}, \bibinfo {author}
  {\bibfnamefont {M.}~\bibnamefont {Reetz-Lamour}}, \ and\ \bibinfo {author}
  {\bibfnamefont {M.}~\bibnamefont {Weidem{\"u}ller}},\ }\href@noop {}
  {\bibfield  {journal} {\bibinfo  {journal} {Phys. Rev. Lett.}\ }\textbf
  {\bibinfo {volume} {99}},\ \bibinfo {pages} {090601} (\bibinfo {year}
  {2007})}\BibitemShut {NoStop}%
\bibitem [{\citenamefont {Schijven}\ \emph {et~al.}(2012)\citenamefont
  {Schijven}, \citenamefont {Kohlberger}, \citenamefont {Blumen},\ and\
  \citenamefont {M{\"u}lken}}]{schijven2012modeling}%
  \BibitemOpen
  \bibfield  {author} {\bibinfo {author} {\bibfnamefont {P.}~\bibnamefont
  {Schijven}}, \bibinfo {author} {\bibfnamefont {J.}~\bibnamefont
  {Kohlberger}}, \bibinfo {author} {\bibfnamefont {A.}~\bibnamefont {Blumen}},
  \ and\ \bibinfo {author} {\bibfnamefont {O.}~\bibnamefont {M{\"u}lken}},\
  }\href@noop {} {\bibfield  {journal} {\bibinfo  {journal} {J.Phys. A}\
  }\textbf {\bibinfo {volume} {45}},\ \bibinfo {pages} {215003} (\bibinfo
  {year} {2012})}\BibitemShut {NoStop}%
\bibitem [{\citenamefont {Schijven}\ and\ \citenamefont
  {M{\"u}lken}(2012)}]{schijven2012avoiding}%
  \BibitemOpen
  \bibfield  {author} {\bibinfo {author} {\bibfnamefont {P.}~\bibnamefont
  {Schijven}}\ and\ \bibinfo {author} {\bibfnamefont {O.}~\bibnamefont
  {M{\"u}lken}},\ }\href@noop {} {\bibfield  {journal} {\bibinfo  {journal}
  {Phys. Rev. E}\ }\textbf {\bibinfo {volume} {85}},\ \bibinfo {pages} {062102}
  (\bibinfo {year} {2012})}\BibitemShut {NoStop}%
\bibitem [{\citenamefont {Gradshteyn}\ and\ \citenamefont
  {Ryzhik}(1980)}]{gradshteyn}%
  \BibitemOpen
  \bibfield  {author} {\bibinfo {author} {\bibfnamefont {I.~S.}\ \bibnamefont
  {Gradshteyn}}\ and\ \bibinfo {author} {\bibfnamefont {I.~M.}\ \bibnamefont
  {Ryzhik}},\ }\href@noop {} {\emph {\bibinfo {title} {Table of Integrals,
  Series, and Products}}}\ (\bibinfo  {publisher} {Academic Press},\ \bibinfo
  {year} {1980})\BibitemShut {NoStop}%
\bibitem [{\citenamefont {Kittel}(1986)}]{kittel}%
  \BibitemOpen
  \bibfield  {author} {\bibinfo {author} {\bibfnamefont {C.}~\bibnamefont
  {Kittel}},\ }\href@noop {} {\emph {\bibinfo {title} {Introduction to Solid
  State Physics}}}\ (\bibinfo  {publisher} {Wiley, New York},\ \bibinfo {year}
  {1986})\BibitemShut {NoStop}%
\bibitem [{\citenamefont {Lahini}\ \emph {et~al.}(2008)\citenamefont {Lahini},
  \citenamefont {Avidan}, \citenamefont {Pozzi}, \citenamefont {Sorel},
  \citenamefont {Morandotti}, \citenamefont {Christodoulides},\ and\
  \citenamefont {Silberberg}}]{lahini2008anderson}%
  \BibitemOpen
  \bibfield  {author} {\bibinfo {author} {\bibfnamefont {Y.}~\bibnamefont
  {Lahini}}, \bibinfo {author} {\bibfnamefont {A.}~\bibnamefont {Avidan}},
  \bibinfo {author} {\bibfnamefont {F.}~\bibnamefont {Pozzi}}, \bibinfo
  {author} {\bibfnamefont {M.}~\bibnamefont {Sorel}}, \bibinfo {author}
  {\bibfnamefont {R.}~\bibnamefont {Morandotti}}, \bibinfo {author}
  {\bibfnamefont {D.~N.}\ \bibnamefont {Christodoulides}}, \ and\ \bibinfo
  {author} {\bibfnamefont {Y.}~\bibnamefont {Silberberg}},\ }\href@noop {}
  {\bibfield  {journal} {\bibinfo  {journal} {Phys. Rev. Lett.}\ }\textbf
  {\bibinfo {volume} {100}},\ \bibinfo {pages} {013906} (\bibinfo {year}
  {2008})}\BibitemShut {NoStop}%
\bibitem [{\citenamefont {Martin}\ \emph {et~al.}(2011)\citenamefont {Martin},
  \citenamefont {Di~Giuseppe}, \citenamefont {Perez-Leija}, \citenamefont
  {Keil}, \citenamefont {Dreisow}, \citenamefont {Heinrich}, \citenamefont
  {Nolte}, \citenamefont {Szameit}, \citenamefont {Abouraddy}, \citenamefont
  {Christodoulides} \emph {et~al.}}]{martin2011anderson}%
  \BibitemOpen
  \bibfield  {author} {\bibinfo {author} {\bibfnamefont {L.}~\bibnamefont
  {Martin}}, \bibinfo {author} {\bibfnamefont {G.}~\bibnamefont {Di~Giuseppe}},
  \bibinfo {author} {\bibfnamefont {A.}~\bibnamefont {Perez-Leija}}, \bibinfo
  {author} {\bibfnamefont {R.}~\bibnamefont {Keil}}, \bibinfo {author}
  {\bibfnamefont {F.}~\bibnamefont {Dreisow}}, \bibinfo {author} {\bibfnamefont
  {M.}~\bibnamefont {Heinrich}}, \bibinfo {author} {\bibfnamefont
  {S.}~\bibnamefont {Nolte}}, \bibinfo {author} {\bibfnamefont
  {A.}~\bibnamefont {Szameit}}, \bibinfo {author} {\bibfnamefont {A.~F.}\
  \bibnamefont {Abouraddy}}, \bibinfo {author} {\bibfnamefont {D.~N.}\
  \bibnamefont {Christodoulides}},  \emph {et~al.},\ }\href@noop {} {\bibfield
  {journal} {\bibinfo  {journal} {Optics Express}\ }\textbf {\bibinfo {volume}
  {19}},\ \bibinfo {pages} {13636} (\bibinfo {year} {2011})}\BibitemShut
  {NoStop}%
\bibitem [{\citenamefont {Segev}\ \emph {et~al.}(2013)\citenamefont {Segev},
  \citenamefont {Silberberg},\ and\ \citenamefont
  {Christodoulides}}]{segev2013anderson}%
  \BibitemOpen
  \bibfield  {author} {\bibinfo {author} {\bibfnamefont {M.}~\bibnamefont
  {Segev}}, \bibinfo {author} {\bibfnamefont {Y.}~\bibnamefont {Silberberg}}, \
  and\ \bibinfo {author} {\bibfnamefont {D.~N.}\ \bibnamefont
  {Christodoulides}},\ }\href@noop {} {\bibfield  {journal} {\bibinfo
  {journal} {Nature Photonics}\ }\textbf {\bibinfo {volume} {7}},\ \bibinfo
  {pages} {197} (\bibinfo {year} {2013})}\BibitemShut {NoStop}%
\end{thebibliography}
\end{document}